\documentclass[twocolumn,preprintnumbers,amsmath,amssymb]{revtex4}

\usepackage{graphicx}
\usepackage{dcolumn}
\usepackage{bm}

\begin{document}


\title{Likelihood based observability analysis and confidence intervals for predictions of dynamic models}

\author{Clemens Kreutz}
 \email{ckreutz@fdm.uni-freiburg.de}
\author{Andreas Raue}%
\author{Jens Timmer}%
\affiliation{%
Freiburg Centre for Biosystems Analysis (ZBSA),\\
Freiburg Institute for Advanced Studies (FRIAS),\\
BIOSS Centre for Biological Signalling Studies,\\
Physics Department,\\
University of Freiburg\\
79104 Freiburg, Germany
}%


\begin{abstract}
Mechanistic dynamic models of biochemical networks such as Ordinary Differential Equations (ODEs) contain unknown parameters like the reaction rate constants and the initial concentrations of the compounds. The large number of parameters as well as their nonlinear impact on the model responses hamper the determination of confidence regions for parameter estimates. At the same time, classical approaches translating the uncertainty of the parameters into confidence intervals for model predictions are hardly feasible.

In this article it is shown that a so-called \emph{prediction profile likelihood} yields reliable confidence intervals for model predictions, despite arbitrarily complex and high-dimensional shapes of the confidence regions for the estimated parameters. 
Prediction confidence intervals of the dynamic states allow a data-based observability analysis. 
The approach renders the issue of sampling a high-dimensional parameter space into evaluating one-dimensional prediction spaces.
The method is also applicable if there are non-identifiable parameters yielding to some insufficiently specified model predictions that can be interpreted as non-observability. 
Moreover, a \emph{validation profile likelihood} is introduced 
that should be applied when noisy validation experiments are to be interpreted.

The properties and applicability of the prediction and validation profile likelihood approaches are demonstrated by two examples, a small and instructive ODE model describing two consecutive reactions, and a realistic ODE model for the MAP kinase signal transduction pathway. 
The presented general approach constitutes a concept for observability analysis and for generating reliable confidence intervals of model predictions, not only, but especially suitable for mathematical models of biological systems.
\end{abstract}

\maketitle

\section{\label{sec:introduction}Introduction}
The major steps of the process of mathematical modeling comprise model discrimination, i.e.~identification of an appropriate model structure, model calibration, i.e.~estimation of unknown model parameters, as well as prediction and model validation. 
For all these topics it is essential to have appropriate methods assessing the certainty or ambiguity of any result for given experimental information.

For parameter estimation, there are several approaches to derive confidence intervals, like standard intervals which are based on an estimate of the covariance matrix of the parameter estimates \cite{sachs84}, bootstrap based confidence intervals \cite{davison97,DiCiccio1987,Joshi2006}, as well as likelihood based confidence intervals \cite{Venzon1988,Raue2009}. 
For model discrimination, significance statements can be obtained by statistical tests. 
However, for model predictions, there are still demands for methodology that is applicable for mathematical models like ODEs used to describe the dynamics of a system in a variety of scientific fields e.g.~in molecular biology \cite{Hlavacek2009,swameye03}, but also in medical research, chemistry, engineering, and physics.

The mere estimation of parameters is often not the final aim of an investigation. More frequently, it is desired to utilize the parametrized model to generate model predictions such as the dynamic behavior of unobserved components. 
Classically, the uncertainty in the model parameters is attempted to be translated into corresponding prediction confidence intervals. For models that depend linearly on the model parameters, as it occurs in classical regression models, 
this is well studied and known as propagation of uncertainty based on standard errors. 
This approach is appropriate and sufficient for many applications. 
However, e.g.~for biochemical networks, the model responses depend nonlinearly on the model parameters. 
Here, the boundaries of the parameter confidence region can exhibit arbitrarily complex shape and are usually difficult to translate into boundaries for the prediction confidence intervals. 
Therefore, established approaches aim to scan the entire parameter subspace which is in a sufficient agreement with the experimental data to propagate the parameters confidence regions into confidence intervals for the model predictions. 
The major challenge is the complex nonlinear interrelation between parameters and model responses which requires that the parameter space has to be densely sampled to capture all scenarios of model predictions. 
For models with tens to hundreds of parameters this is numerically demanding or even infeasible because high dimensional spaces cannot be densely sampled. 
This is issue often referred to the curse of dimensionality in literature \cite{Marimont79,Scott1991}. 

There are several methods for an approximate sampling of the parameter space, e.g.~the Markov Chain Monte Carlo (MCMC) methods \cite{Gelman03,kass98}, 
and bootstrap based approaches \cite{Joshi2006,Molinaro2005}. 
However, for the ODE models used to describe interaction networks, these methods are numerically very demanding and provide only very rough approximations. 
Therefore, it is difficult to control the coverage of the prediction confidence intervals for these approaches. 
Moreover non-identifiable parameters are not explicitly considered hampering the convergence of these sampling techniques and yielding results that are questionable and difficult to interpret \cite{Bayarri:2004}.

Conceptually related to the prediction profile likelihood approach presented here, \cite{Raue2009, Raue:2010fk} presented an approach for the determination of confidence intervals for the model parameters by sampling the parameter profile likelihood. 
For their approach the problems concerning identifiability are resolved and the computation is feasible also for high-dimensional parameter spaces. 
Nevertheless, the direct translation to confidence intervals of the model trajectories only works as an approximation yielding a coverage rate that is sometimes lower than desired.

The idea of the prediction profile likelihood presented here is to determine prediction confidence without an explicit sampling strategy for the parameter space. 
Instead, a certain fixed value for a prediction is used as a nonlinear constraint and the parameter values are chosen 
via constraint optimization of the likelihood.
This does neither require a unique solution in terms of identifiability nor confidence intervals for the parameter estimates. 
The constraint maximum likelihood approach checks the agreement of a predicted value with the experimental data. 
By repeating this procedure for continuous variations of the predicted value, a \emph{prediction profile-likelihood} is obtained.
Thresholding the prediction profile likelihood yields statistically accurate confidence intervals.
The desired level of confidence which coincides with the level of agreement with the experiments is controlled by the threshold.

The theoretical background of the prediction profile likelihood, also called \emph{predictive likelihood} has been already studied \cite{Hinkley79}.
Moreover, related ideas are already applied in the context of generalized linear mixed models \cite{Booth98},
unobserved data points \cite{Butler86}. 
The linear approximation has been applied in nonlinear regression analyses \cite{Cooley89}.
A review of prediction profile likelihood approaches and a modification to sufficiency-based predictive likelihood is provided in \cite{Bjornstad1990}.

In this paper, this concept is applied to ODE models occurring in mechanistic dynamic models, e.g.~in molecular and systems biology.
In this context the approach allows a data-based observability analysis.
Moreover, it is extended to obtain confidence intervals for validation experiments.

\section{\label{sec:level2}Results}
\begin{figure*}
\begin{center}
\includegraphics[width=\linewidth,height=12cm]{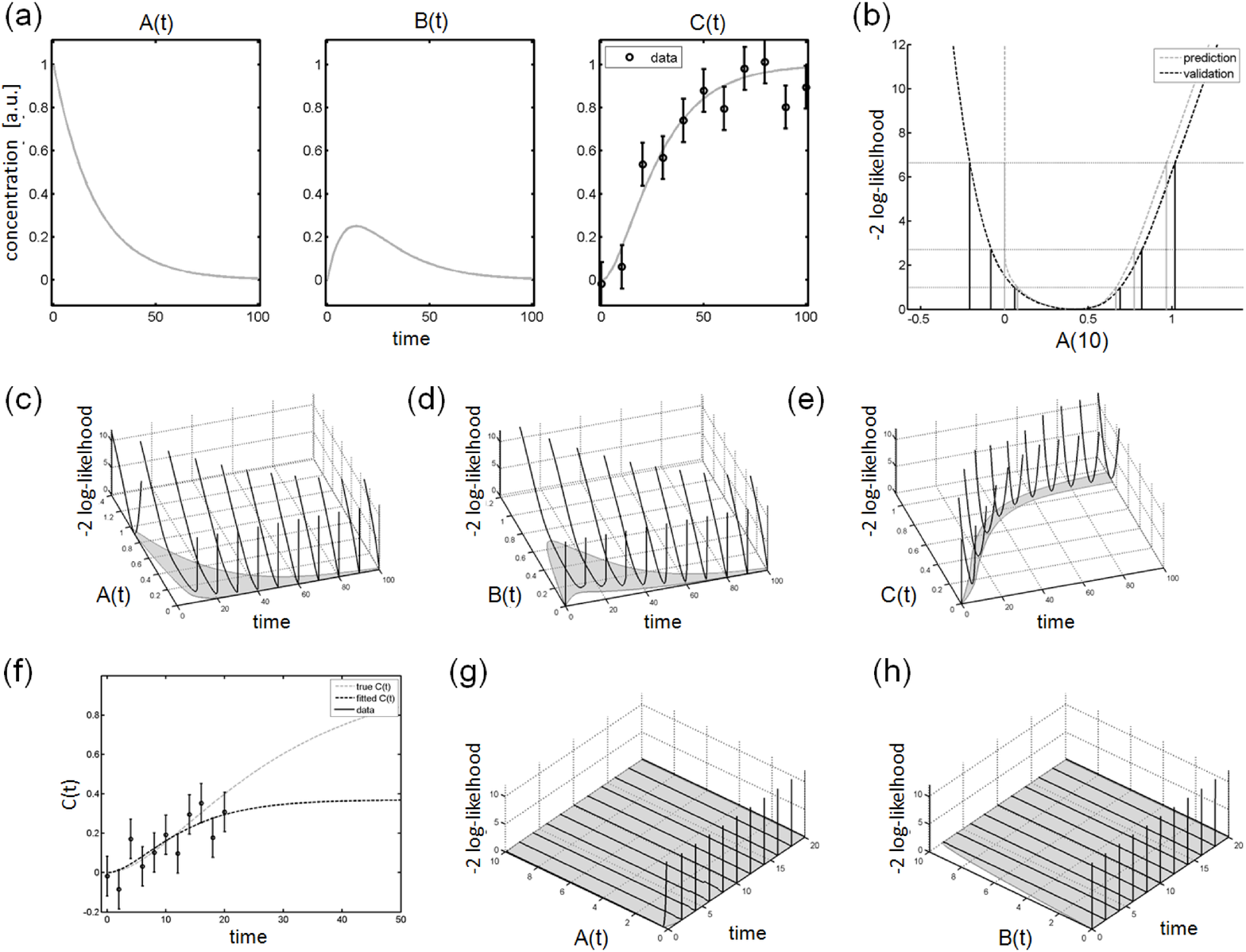}
\end{center}
\caption{
The three figures in panel (a) show the dynamics and measurement realization for the small model used for illustration purpose. 
C(t) is measured and the dynamics of all states, i.e.~A(t), B(t), and C(t), is intended to be predicted.
Panel (b) shows as an example the prediction profile likelihood (gray dashed curve) and validation profile likelihood (black dashed curve) of A(t=10).
Thresholding yields confidence intervals for prediction (gray vertical lines) and validation (black vertical lines).
The three thresholds and the respective projections correspond to $\alpha$=68\%, 90\%, and 99\% confidence intervals.
The VCIs are larger than the PCIs, because they account for the measurement error of a validation data point.
Panels (c)-(e) show prediction confidence intervals (gray) for the unobserved states A(t), B(t),
as well as for the measured state C(t).
The prediction profile likelihood functions are plotted as black curves in vertical direction.
Non-observability is illustrated in panels (f)-(h). 
Panel (f) shows a realization of the measurements for a design which does not provide sufficient information
about the steady state of C.
This leads to a flat prediction profile likelihood for large values for A(t) as shown in panel (g),
as well as for B(t) for t$>$0 as plotted in panel (h).
A flat prediction profile likelihood in turn yields unbounded prediction and validation confidence intervals and non-observability of A(t) and B(t) as indicated by the gray shaded regions.
\label{fig:cons}
}
\end{figure*}
\subsection{Small illustration model}
First, a small but illustrative model of two consecutive reactions
\begin{equation}
     A\: \overset{\theta_1}{\rightarrow} \:B\: \overset{\theta_2}{\rightarrow} \:C
\end{equation}
with rates $\theta_1=0.05, \theta_2=0.1$ and initial conditions $A(0)=\theta_3=1, B(0)=0, C(0)=0$ is utilized to illustrate our approach. 
For this purpose, it is assumed that $C(t)$ is measured at $t=0,10,\dots,100$.

For the simulated measurements, Gaussian noise $\varepsilon \sim N(0,\sigma^2)$ with $\sigma=0.1$ has been assumed which corresponds to a typical signal-to-noise ratio for applications in cell biology of around $10\%$. 
If an experimental setup would not allow for negative measurements, a log-normal distribution of the observational noise could be more appropriate.
Then, the Gaussian setting is obtained after a log-transformation of the data \cite{Kreutz2007}.
Such transformations and preprocessing procedures would have to be performed before the analysis starts.

Panel (a) in Fig.~\ref{fig:cons} shows the dynamics of $A(t)$, $B(t)$, and $C(t)$ as well as a typical noise realization.
Such simulated data realizations are utilized to calculate the prediction- and validation profile likelihood for the 
dynamic states. 

Panel (b) shows, as an example, the prediction profile likelihood and the validation profile likelihood
for the same noise realization for predicting $A(t)$ at time point $t=10$.
The validation profile likelihood has been calculated for validation data with 10\% measurement noise, 
as it was assumed for the measurements
The vertical axis is minus two times the log-likelihood which corresponds to the residual sum of squares. 
For illustration purposes, the minimum of the log-likelihood $\text{LL}^*$ is shifted to zero in all figures.
Three thresholds corresponding to 68\%, 90\%, and 99\% confidence levels are plotted as horizontal lines.
As explained in the section `Materials and Methods', the projections to the horizontal axis yields the respective confidence intervals for a prediction or for a validation experiment.
The constraint optimization procedure is infeasible for $A(t)\le 0$ and therefore the PCIs automatically account for strictly positive values of $A$.

The calculation of the prediction and validation confidence intervals has been repeated for $t=0,10,\dots,100$ and all three dynamic states $A(t)$, $B(t)$, $C(t)$.
In panels (c)-(e), the respective prediction confidence intervals (PCIs) are plotted as well as the prediction
profile likelihood.
The corresponding validation profile likelihood functions and the respective validation confidence intervals are shown in the supplemental information (SI).
For plotting the confidence intervals along the time axis, the $\text{PCIs}$ evaluated the eleven time points have been interconnected by cubic piecewise interpolation.
The displayed confidence intervals constitute the propagation of information from the measurements of $C(t)$ to predictions of the dynamics of the compound concentrations.
Because $C$ is the measured compound in our example, the prediction confidence intervals for $C$ are much smaller than for $A$ and $B$.
However, also $A$ and $B$ yield bounded prediction confidence intervals which can be interpreted as {\it observability} of these dynamic states.

In the Appendix, the reliability of our confidence intervals is investigated by calculating the coverage, 
i.e.~by comparing the confidence level with the frequency that the true value is inside the prediction confidence interval.
For our example, it will be demonstrated that confidence intervals 
using the asymptotic threshold sometimes yield slightly conservative intervals.
Also an algorithm to improve the threshold is provided which yields slightly smaller confidence intervals with the correct coverage.

To illustrate the effect of {\it non-observability}, the assumption about the available experimental information 
is slightly changed.
The measurements are simulated for earlier and closer time steps, i.e.~for $t=0,2,\dots,20$.
Panel (f) of Fig.~\ref{fig:cons} shows that these time points sample only the transient increase of $C(t)$.
Hence, such a design does not provide sufficient information about the steady state level of $C$.
In other words, the modification limits the available information about the total amounts of the compounds.
This, in turn, makes $A(t)$ and $B(t)$ non-observable.

Panel (g) shows the prediction confidence intervals for $A(t)$.
In the chosen setting, the predictions are unbounded towards infinity and therefore $A(t)$ is non-observable.
In panel (h), it is also shown that $B(t)$ is non-observable.
According to the model definition, $B(0)$ is known to be zero, but for
$t>0$, unbounded prediction confidence intervals are obtained which indicate non-observability of $B(t)$.

\subsection{MAP kinase signaling model}
Next, an ODE model of cellular signal transduction has been used to illustrate our method in a realistic setting.
For this purpose, a model of the \emph{mitogen-activated protein (MAP) kinases} which is one of the most extensively studied signal transduction pathway, is utilized. 
The chosen model \cite{Kholodenko2000} consists of eight dynamic states describing the time dependency of the MAP kinases Raf, Raf$^*$, Mek, Mek$^*$, Mek$^{**}$, Erk, Erk$^*$, and Erk$^{**}$ which play a very prominent role in many cellular processes, e.g.~in cell proliferation.
A star `*' denotes phosphorylation of the protein which biologically acts as activation.

\begin{figure*}
\begin{center}
\includegraphics[width=\linewidth,height=5cm]{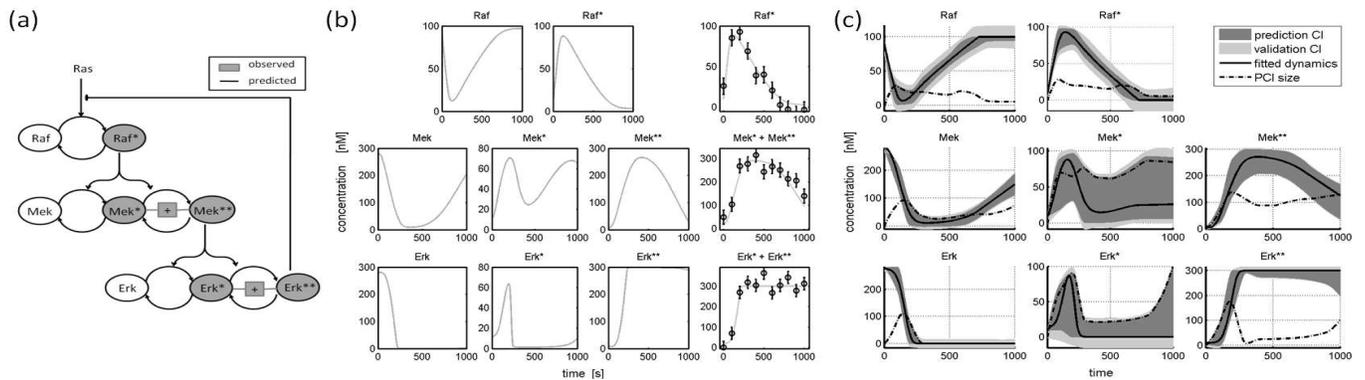}
\end{center}
\caption{Panel (a) shows the MAP kinase model according to \cite{Kholodenko2000}. 
It is assumed that the phosphorylated compounds are measured.
The dynamics of all compounds is intended to be predicted to illustrate the prediction profile likelihood approach. 
In panel (b) the dynamics of the MAP kinase model
as well as simulated data set are plotted.
The 90\% confidence intervals of the dynamic variables for predictions (dark gray) and for validation experiments (light gray) for this noise realization are plotted in panel (c). 
The size of the PCI is plotted as a dashed-dotted line. 
In absolute concentrations, the dynamics of Erk$^{**}$ has the largest PCI at t=181 seconds, i.e.~when the negative feedback is activated.
Also, the dynamics of Mek$^*$ is only badly observable in our example.
Measurements of both would be very informative for better calibrating the model.
\label{fig:kholo}}
\end{figure*}
Panel (a) in Fig.~\ref{fig:kholo} provides a summary of the MAP kinase signaling pathway.
The enzymatic reactions in the ODE model are described as Michaelis-Menten rate equations,
i.e.~each reaction is parametrized with a maximal enzymatic rate and a Michaelis constant.
As in the original publication, the parameters of the two consecutive phosphorylation and dephosphorylation steps of Mek and Erk are assumed to be identical and the initial concentrations are assumed to be known.
In this setting, 14 parameters are estimated out of three times eleven data points.
Details about the model are provided in the Appendix.

It is assumed that the total amount of the phosphorylated forms for each protein, i.e.~Raf$^{*}$, the sum of Mek$^{*}$ and Mek$^{**}$ as well as the sum of Erk$^{*}$ and Erk$^{**}$, are measured.
This observational assumption holds for example for phospho-specific antibodies such as utilized for Western blotting.
The measurement times are set to $0,100,\dots,1000$ seconds.
Again, additive Gaussian noise is assumed. 
The standard deviation has been set to $\sigma=10$ nM.

In panel (b) of Fig.~\ref{fig:kholo} a typical noise realization is displayed. 
Panel (c) shows the prediction confidence intervals (dark gray) and the validation confidence intervals (light gray) for this noise realization calculated for all dynamic states.
The size of the confidence intervals is plotted as a dashed-dotted line.

The prediction confidence intervals show how precisely the dynamics is inferred by the available data.
The temporal behavior of Raf, Raf$^*$ is quite well determined, i.e.~the size of the PCI is below 40 nM. 
Similarly, the unphosphorylated states of Mek and Erk have narrow prediction confidence intervals. 
For Mek$^*$ the concentration dynamics is only predicted within rather large intervals which for most time points 
nearly span a range between zero and 100 nM.

The largest absolute size of the prediction confidence interval of 176 nM is obtained for Erk$^{**}$ after 181 seconds.
This is the point in time where the negative feedback is activated.
Additional experimental investigation of this condition is very informative to further specify the dynamic behavior of the MAP kinase cascade in our example. 
Further considerations concerning experimental planning are provided in detail in the Appendix.

\section{Discussion}
Existing approaches for prediction confidence intervals like MCMC \cite{Marjoram} or bootstrap procedures are based on forward evaluations of the model for many parameter values. 
This works reasonably well for a low dimensional parameter space and if the target density function, i.e.~the parameter space to be sampled, is well-behaved \cite{Bayarri:2004}. 
However, sampling nonlinear high-dimensional spaces densely is impractical and it is almost impossible to ensure that sampling the parameter space covers all prediction scenarios.
Especially in biological applications the target distributions frequently inherit strong and nonlinear functional relations. 
In the case of non-identifiability, the parameter space to be sampled is not restricted 
rendering convergence near to impossible.

In this paper, we applied a contrary procedure.
The model prediction space is sampled directly and the corresponding model parameters are determined by constraint maximum likelihood to check the agreement of the predictions with the data.
This concept yields the prediction profile likelihood which constitutes the propagation of uncertainty from experimental data to predictions.

If a comprehensive prior, i.e.~for all parameters, would be available, a Bayesian procedure like MCMC where marginalization, i.e.~integration over the nuisance dimensions is feasible could have superior performance.
However, in cell biology applications, prior knowledge is very restricted because kinetic rates and concentrations are highly dependent on the cell type and biological context, e.g.~on the cellular environment and biochemical state of a cell.
Therefore, there is usually at most some prior information for few parameters available.
Such prior information can be incorporated in our procedure without restricting its applicability by generalizing maximum likelihood estimation to maximum a-posterior estimation as discussed in the Appendix.

In general, generating prediction confidence intervals given the uncertainty in the high-dimensional nonlinear parameter space requires large numerical efforts.
However, this complication primarily originates from the complexity of the issue itself rather than from the methodological choice. 
In fact, the aim is approached by the prediction profile likelihood in a very efficient manner because 
scanning the parameter space by the constrained optimization procedure to explore the data-consistent predictions is more efficient than sampling parameter space without considering the predictions like it is performed for MCMC.
Instead of sampling a high-dimensional parameters space, only the prediction space has to be explored for calculating a prediction profile likelihood, i.e.~the optimization of the parameters reduces the high-dimensional sampling problem to exploring a single dimension.
For the parameter profile likelihood, it has been demonstrated \cite{Raue2009} that the computational effort only scales slightly super-linear with the number of parameters. 
This result does, due to the similarity of the computations, carry over to the prediction profile likelihood.

The prediction confidence regions introduced above has to be interpreted \emph{point-wise}. 
This means that a confidence level $\alpha$ controls errors of type 1
which is the probability that the model response for the true parameters is inside the prediction confidence interval 
for a single prediction condition if many realizations of the experimental data and the corresponding prediction confidence intervals are considered.

In contrast, if a single data set is utilized to generate many prediction intervals, 
e.g.~predictions for several points in time as performed above, the results are \emph{statistically dependent}, 
i.e.~the realization of the $\text{PCI}$ of a neighboring time point is very similar and therefore correlated. 
Therefore, the prediction confidence intervals for a compound for two adjacent points in time very likely both contain the true value, or neither.
In such an example, a \emph{common} prediction confidence region for two statistically dependent predictions would require a two-dimensional prediction profile likelihood.
This topic, however, is beyond the scope of this article.


The prediction profile likelihood also provides a concept for experimental planning. 
Experimental conditions with a very narrow prediction confidence interval are very accurately specified by the available data. 
New measurements for such a condition on the one hand does not provide very much additional information to better calibrate the model parameters, and hence is from this point of view a bad choice for additional measurements. 
On the other hand, it very precisely predicts the model behavior under these certain conditions and is therefore a very powerful candidate setting for validating the model structure.
Contrarily, large prediction confidence intervals indicate conditions which are weakly specified by the existing data
and therefore constitute informative experimental designs for better calibrating the model.
Because a design optimization on the basis of the prediction profile likelihood does not require any linearity approximation like common experimental design techniques, e.g.~based on the \emph{Fisher information} \cite{Kreutz2009}, the presented procedure could be very valuable for ODE models which are typically highly nonlinear.

Another potential of the prediction profile likelihood shown in this article is its interpretation in terms of \emph{observability}. 
This term is very commonly used in control theory to characterize whether the dynamics of some unobserved variables can be inferred by the set of feasible experiments.
The theory in this field is based on analytical calculations, i.e.~the limited amount and inaccuracy of the data 
is usually not considered.
In this article, it has been shown that the prediction profile likelihood allows for a general data-based approach  to check whether there is enough information about unobserved  dynamic states in the given experimental design and realization of measurements. 
Therefore, in analogy to the terminology of {\it practical identifiability} \cite{Raue2009}, we would suggest to term observability for a given data set, i.e.~a restricted prediction confidence interval, as \emph{practical observability}.


Finally, it should be noted, that a prediction could be any function of the compounds and the parameters.
In applications, e.g.~a ratio of two compound concentrations is a characteristics of interest.
In principle also integrals, peak positions and other functions of the dynamic states can be considered as predictions
which could be targets for observability considerations as well as for the calculation of prediction and validation confidence intervals.

\section{Summary}
Generating model predictions is a major task in mathematical modeling.
For the dynamic mechanistic models as they are applied e.g.~in molecular and systems biology, the confidence regions from parameter estimation can have arbitrarily complex shapes. Therefore, it is very difficult or even impossible to sample the parameter space appropriately to generate confidence intervals for predictions. 
This in turn impedes a data-based observability analysis for the dynamic states.

In this article, the prediction profile likelihood approach is presented as a methodology which directly calculates the set of model predictions which are consistent with existing measurements. 
This concept constitutes a powerful tool for assessing model predictions, performing observability analyses, and experimental design.
The method is feasible for arbitrary dimensions of the parameter space. 
It only requires a proper calculation of the maximum likelihood value, 
i.e.~a numerically working parameter optimization procedure.
The task of sampling a high-dimensional parameter space reduces to scanning a one-dimensional prediction space.
It therefore allows the calculation of confidence intervals for model predictions as well as 
confidence intervals for the outcome of validation experiments.

The applicability of the approach has been shown by a small but instructive system of two consecutive reactions
and a published model for MAP kinase signaling. 
For the small system, 
it has been shown that the prediction profile likelihood yields desired coverage properties. 
Moreover, a setting inducing non-observability has been investigated which is characterized by unbounded prediction confidence intervals.
For the MAP kinase model, prediction confidence intervals and validation confidence intervals for all dynamic states have been determined on the basis of measurements of the phosphorylated proteins.
In addition, the applicability of the approach for experimental planning has been demonstrated.


\section{Methodology}
\subsection{The prediction profile likelihood}
For additive Gaussian noise $\varepsilon \sim N(0,\sigma^2)$ with known variance $\sigma^2$, two times the negative log-likelihood
\begin{equation}
	-2\,\text{LL}(y|\theta) = \sum_i \frac{\left(y_i - F(t_i,u,\theta)\right)^2}{\sigma^2} + const.  \label{eq:LL}
\end{equation}
of the data $y$ for the parameters $\theta$ is except a constant offset identical to the residual sum of squares
$	\text{RSS}(\theta) =  \sum_i \left( y_i - F(t_i,u,\theta)\right)^2 /\sigma^2$.
In this case, maximum likelihood estimation is equivalent to standard least-squares estimation
\begin{equation}
	\hat \theta = \arg \max_{\theta} \: \text{LL}(y|\theta) \equiv  \arg \min_{\theta} \: \text{RSS}(\theta)\:,
\end{equation}
i.e.~to minimizing the residual sum of squares.
$F=g(x(t,u,\theta),\theta)$ denotes the model response which is in our case given after integration of a system of differential equations
\begin{equation}
	\dot x(t) = f(x(t), u(t), \theta)  \label{eq:ODE}
\end{equation}
with an externally controlled input function $u$ and a mapping to experimentally observable quantities
\begin{equation}
	y(t) = g(x(t), \theta) + \varepsilon(t) \label{eq:OBS}.
\end{equation}
The parameter vector $\theta$ comprises the kinetic parameters as well as the initial values, 
and additional offset or scaling parameters for the observations.\\
It has been shown \cite{Raue2009} that the profile likelihood 
\begin{equation}
	\text{PL}(\theta_i)  = \max_{\theta_{j\ne i}}  \text{LL}(\theta|y)			\label{eq:PL}			
\end{equation}
for a parameter $\theta_i$ given a data set $y$ yields reliable confidence intervals 
\begin{equation}
	\text{CI}_{\alpha}(\theta_i|y) = \left\{\theta_i \:|\: -2\text{PL}(\theta_i) \le -2\text{LL(y)}^* +  icdf(\chi^2_1,\alpha) \right\}	\label{eq:CI}
\end{equation}
for the estimation of a single parameter. 
Here, $\alpha$ is the confidence level and $icdf(\chi^2_1,\alpha)$ denotes the $\alpha$ quantile of the chi-square distribution with one degree of freedom which is given by the respective inverse cumulative density function. 
$\text{LL}^*$ is the maximum of the log-likelihood function after all parameters are optimized.
In [\ref{eq:PL}], the optimization is performed for all parameters except $\theta_i$. 
The analogy of likelihood-based parameter and prediction confidence intervals is discussed in the Appendix.
\\
The desired coverage
\begin{equation}
	\text{Prob}\left(\theta_i \in \text{CI}_{\alpha}(\theta_i) \right) = \alpha \label{eq:coverageP} \:\:,
\end{equation}
i.e.~the probability that the true parameter value is inside the confidence interval, 
holds for [\ref{eq:CI}] if the magnitude of the decrease of the residual sum of squares by fitting of $\theta_i$ is $\chi^2_1$ distributed. 
This is given asymptotically as well as for linear parameters and is a good approximation
under weak assumptions \cite{Feder1968,seber89}. 
If the assumptions are violated, the distribution of the magnitude of the decrease 
has to be generated empirically, i.e.~by Monte-Carlo simulations, as discussed in the Appendix.
\\
The \emph{experimental design} $D=\{t,g,u\}$ comprises all environmental conditions which can be controlled by the experimenter like the measurement times $t$, the observables $g$, and the input functions $u$.
A \emph{prediction} $z = F(D_{\text{pred}},\theta)$ is the response of the model $F$ for a prediction condition $D_{\text{pred}}=\{t_{\text{pred}},g_{\text{pred}},u_{\text{pred}}\}$ specifying a prediction observable $g_{\text{pred}}$ evaluated at time point $t_{\text{pred}}$ given the externally controlled stimulation $u_{\text{pred}}$. 
\\
In some cases the observable $g_{\text{pred}}$ corresponds to measuring a dynamic variable $x(t)$ directly, i.e.~it corresponds to a compound whose concentration dynamics is modeled by the ODEs. 
In a more general setting the observable is defined by an observational function $g_{\text{pred}}(x(t), \theta)$ depending on several dynamic variables $x$.
Therefore, $g_{\text{pred}}$ does neither have to coincide with a dynamic variable nor with an observational function $g$ of the measurements performed to build the model. 
\\
In analogy to [$\ref{eq:coverageP}$], the desired property of a prediction confidence interval $\text{PCI}_{\alpha}(D|y)$ derived from an experimental data set y with a given significance level $\alpha$ is that the probability 
\begin{equation}
	\text{Prob}( F(D_{\text{pred}},\theta_{\text{true}}) \in \text{PCI}_{\alpha}(D|y) ) = \alpha   \label{eq:coverage}
\end{equation}
that the true value of $F(D_{\text{pred}},\theta_{\text{true}})$ is inside the prediction confidence interval $\text{PCI}_{\alpha}$ is equal to $\alpha$. In other words, the $\text{PCI}$ covers the model response for the true parameters with a proportion $\alpha$ of the noise realizations which would yield different data sets $y$.  
\\
The prediction profile likelihood 
\begin{equation}
	\text{PPL}(z) = \max_{ \theta \in \{\theta | F(D_{\text{pred}},\theta)=z\}}  \: \text{LL}(y|\theta)  \label{eq:optiPPL}
\end{equation}
is obtained by maximization over the model parameters satisfying the constraint that the model response $F(D,\theta^*)$ after fitting is equals to the considered value $z$ for the prediction.
The prediction confidence interval is in analogy to [$\ref{eq:CI}$] given by
\begin{equation}
\begin{small}
	\text{PCI}_{\alpha}(D_{\text{pred}}|y) = \left\{z \,|\, -2\text{PPL}(z) \le -2\text{LL}^*(y) + icdf(\chi^2_1,\alpha) \right\} \:,	\label{eq:PCI}
\end{small}
\end{equation}
i.e.~the set of predictions $z=F(D_{\text{pred}},\theta)$ for which the $\text{PPL}$ is below a threshold given by the $\chi^2_1$ distribution.
In analogy to likelihood based confidence intervals for parameters, such PCI yields the smallest unbiased confidence intervals for predictions for given coverage $\alpha$ \cite{cox94}.
\\
Instead of sampling a high-dimensional parameter space,
the prediction profile likelihood calculation comprises sampling of a one-dimensional prediction space 
by evaluating several predictions $z$.
Evaluating the maximum of the likelihood satisfying the prediction constraint does in general not require an unambiguous point in the parameter space as in the case of structural non-identifiabilities.
In analogy to profile likelihood for parameter estimates, the significance level determines the threshold for the $\text{PPL}$, which is given asymptotically by the quantiles [$\ref{eq:CI}$] of the $\chi^2_1$ distribution \cite{Meeker}.
In the Appendix, a Monte-Carlo algorithm is presented which can be used to calculate the threshold in cases where the asymptotic assumption is violated.

\subsection{The validation profile likelihood}
Likelihood-based confidence interval like [$\ref{eq:CI}$] or [$\ref{eq:PCI}$] correspond to the region 
where a likelihood ratio test would not reject the model. 
Having a prediction confidence interval, the question arises whether a model has to be rejected if a validation measurement is outside the predicted interval. 
This, in fact, would hold if a ``perfect'' validation measurement would be available, i.e.~a data point without measurement noise. 
For validation experiments, however, the outcome is always noisy 
and is therefore expected to be more frequently outside the \text{PCI} than the true value.
Hence, the prediction confidence interval [$\ref{eq:PCI}$] has to be generalized for application to a validation experiment.
\\
For a validation experiment, we therefore introduce a \emph{validation profile likelihood} $\text{VPL}$ and a corresponding
\emph{validation confidence interval} $\text{VCI}^{\text{SD}}_{\alpha}$ in the following.
In such a setting, a confidence interval should have a coverage
\begin{equation}
	\text{Prob}\left( z  \in \text{VCI}_{\text{SD}}^{\alpha}( D_{\text{vali}}|y) \right) = \alpha   \label{eq:coverageNoisy}
\end{equation}
for the validation data point $z \sim N(\mu,\text{SD}^2)$ with
expectation $\mu = F(D_{\text{vali}},\theta_{\text{true}})$ and variance $\text{SD}^2$. 
Here, $D_{\text{vali}}$ denotes the design for the validation experiment.
A validation confidence interval satisfying [$\ref{eq:coverageNoisy}$] allows a rejection of the model if a noisy validation measurement with error $\text{SD}$ is outside the interval.
\\
$\text{VCI}_{\alpha}^{\text{SD}}$ for validation data can be calculated by relaxing the constraint [$\ref{eq:optiPPL}$] used to compute the prediction profile likelihood.
Because in this case, the model prediction does not necessarily have to coincide with the data point $z$. 
Instead, the deviation from the validation data point is penalized equivalently to the data $y$. 
The agreement of the model with the data $y$ and the validation measurement $z$ is then given by
\begin{equation}\begin{small}
\text{LL}(z,y|\theta)  
	= \sum_i \left ( \frac{y_i - F(D_i,\theta)}{\sigma} \right)^2   
	+ \left( \frac{z - F(D_{\text{vali}},\theta)}{\text{SD}}  \right)^2 \label{eq:llNoisy}
\end{small}
\end{equation}
We now define the validation profile (log-)likelihood 
\begin{equation}
	\text{VPL}^{\text{SD}}(z|y) = \text{LL}^*(z,y) = \text{LL}(z,y|\theta^*)
\end{equation}
with 
$	\theta^* = \theta^*(z,y) = \arg \max_\theta \: \text{LL}(z,y|\theta)$
as the maximized joint log-likelihood in [$\ref{eq:llNoisy}$] read as a function of $z$.
The corresponding validation confidence interval is given by
\begin{equation}\begin{small}
	\text{VCI}_{\alpha}^{\text{SD}} (D_{\text{vali}}|y) = \left\{z | -2\text{VPL}^{\text{SD}}(z|y) \le -2\text{LL}^*(z,y) + icdf(\chi^2_1,\alpha) \right\}.
\end{small}
\end{equation}
\\
Optimization of the likelihood [$\ref{eq:llNoisy}$] minimizes both, the contribution of the data $\text{RSS}(y)$, and the mismatch with the fixed prediction value $z$. 
The model response $F(D_{\text{pred},\theta^*})$ obtained after this parameter optimization can be interpreted as
a prediction $z'$ satisfying the constraint optimization problem [$\ref{eq:optiPPL}$]
considered for the prediction profile likelihood.
It holds
\begin{equation}
\begin{small}
	\text{LL}^*(z,y; \text{SD}>0) - \frac 1 2 \frac{\left(z-F(D_{\text{vali}},\theta^*) \right)^2}{\text{SD}^2}= \text{LL}^*(z',y; \text{SD}=0 )  \:\:,
\end{small}
\end{equation}
i.e.~the validation profile likelihood $\text{LL}^*$ can be scaled to the prediction likelihood via
\begin{equation}
	\text{PPL}(z'|y) = \text{VPL}^{\text{SD}}(z|y) - \frac 1 2 \frac{(z'-z)^2}{\text{SD}^2}  \label{eq:pplRescaling}
\end{equation}
where $z'= F(\theta^*(z,y,\text{SD}>0))$ is the model response for $\theta^*$ estimated from $z$ and $y$.
\\
Optimization with nonlinear constraints is a numerically challenging issue. 
Therefore,  [$\ref{eq:pplRescaling}$] provides a helpful way to omit constraint optimization. The $\text{VPL}$ can be calculated with $\text{SD}>0$ like a common least-squares minimization and is then afterwards rescaled to obtain the $\text{PCI}$ for the true value.



\begin{acknowledgments}
The authors thank our long-term experimental collaboration partners, especially Dr.~Maria Bartolome-Rodriguez and Prof.~Ursula Klingm\"uller and their groups for their support and their experience in practically relevant issues.
In addition, the authors acknowledge financial support provided by the BMBF-grants 0315766-VirtualLiver, 0315415E-LungSys and 0313921-FRISYS.
\end{acknowledgments}

\newpage.\newpage
\appendix
\section{Supplementary information}
\subsection{Re-parametrization}
Parameter estimation, i.e.~the prediction of a parameter value out of experimental data, 
can be seen as a special case of a model prediction.
Then, the parameter profile likelihood coincides with the prediction profile likelihood and the respective parameter confidence intervals correspond to prediction confidence intervals.
In this sense, the prediction profile likelihood generalizes the parameter profile likelihood.
In fact, the idea of the prediction profile likelihood and the calculation of prediction confidence intervals, 
e.g.~the choice of the threshold,
is very intuitive for this special case. 

In other situations, an analog strategy would require a re-parametrization of the model in a way that the desired model prediction is unambiguously given by the value of a single new parameter.
Then, again the profile likelihood for the new parameter would give a confidence interval for the prediction. 
In this case, without loss of generality such a parameter can be denoted by $\theta_1'$. 
Then, the re-parametrization would be a transformation 
\begin{equation}
	T: \{\theta_1,\dots,\theta_{n_p}\} \rightarrow \{ \theta'_1,\dots,\theta'_{n_p} \}  \label{eq:repara}
\end{equation}
of the $n_p$ parameters $\theta$ to new parameters $\theta'_{1},\dots,\theta'_{n_p}$ where all predictions for the condition $D_{\text{pred}}$ satisfy
\begin{equation}
	F'(D_{\text{pred}},\theta')  = F'(D_{\text{pred}},\theta_1') \label{eq:T} \:\:.
\end{equation}
Here, $F'= F \circ T^{-1}$ denotes the model for the transformed parameters. 
For a transformation satisfying [$\ref{eq:T}$], any change of the parameters $\theta'_2,\dots,\theta'_{n_p}$ would not affect $F'$, because $T$ is chosen in a way that the effect of $\theta_2, \dots, \theta_{n_{\theta}}$ is orthogonal to the effects of $\theta_1$. 


However, because ODE systems can only be solved analytically for special cases, such a re-parametrization cannot be found explicitly for most realistic models. 
This restriction can be resolved numerically by an implicit re-parametrization which is obtained by a constrained nonlinear optimization procedure. This idea yields the prediction profile likelihood 
\begin{equation}
	\text{PPL}(z) = \max_{ \theta \in \{\theta | F(D_{\text{pred}},\theta)=z\}}  \: \text{LL}(y|\theta)  \label{eq:optiPPL2}
\end{equation}
which is obtained by maximization over the model parameters satisfying the constraint that the model response $F(D,\theta^*)$ after fitting is equals to the considered value $z$ for the prediction.
In this case, the `new parameter' is the predicted value itself, i.e. $z\equiv \theta_1'$ and $F'$ is the identity function.

Equation [$\ref{eq:optiPPL2}$] also resolves the formal issue which occurs if there is not a unique parameter set $\theta$ given by the constraint $F(D_{\text{pred}},\theta)=z$. 
If there are several such parameter sets, the ambiguities either vanish by taking the parameter set 
with maximize the log-likelihood, or they are not relevant because only the value of the maximized log-likelihood enters the calculation.

\subsection{Profile likelihood threshold}
\begin{figure}
\begin{center}
\includegraphics[width=\linewidth]{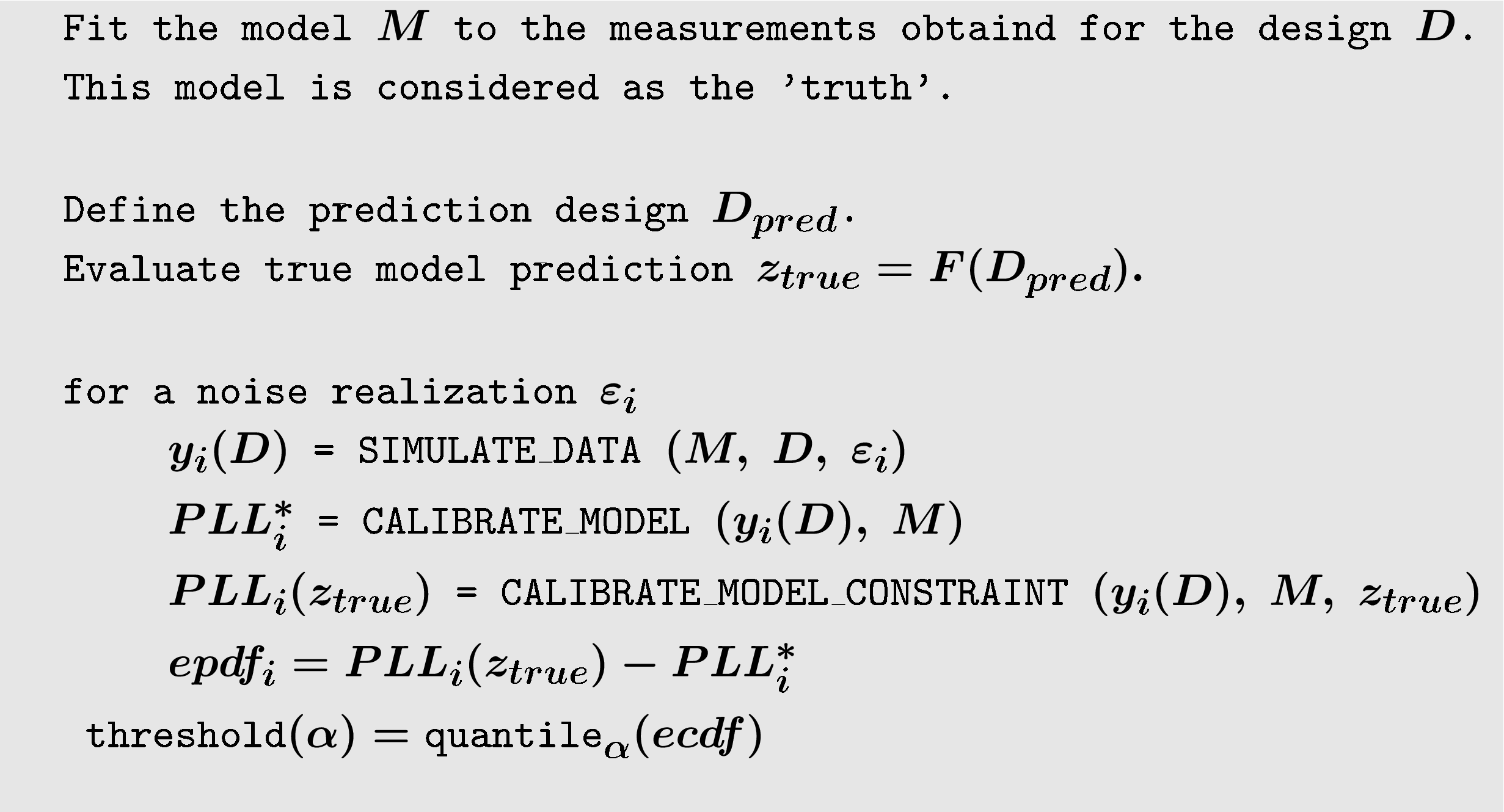}
\end{center}
\caption{A Monte-Carlo algorithm for calculating the profile likelihood threshold empirically.
New noise realizations $y_i(D)$ are utilized to calculate the distribution of $\text{PPL}_i(z_{\text{true}})-\text{PPL}_i^*$.
The $\alpha$ quantile of this distribution can be used as a threshold for prediction confidence intervals instead of the 
asymptotic threshold, i.e.~instead of the $\alpha$ quantile of the $\chi^2_1$ distribution.
\label{fig:checkThreshold}}
\end{figure}
\begin{figure}
\begin{center}
\includegraphics[width=\linewidth]{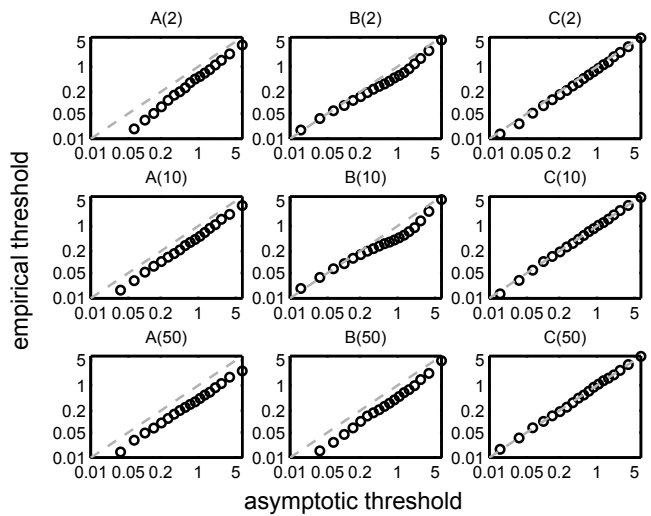}
\end{center}
\caption{The Monte-Carlo approach allows a comparison of the asymptotic thresholds with the empirically calculated, i.e.~the correct thresholds.
Here, the thresholds corresponding to $0.05,0.1,\dots,0.95,0.99$ confidence levels have been plotted for nine different prediction scenarios. 
In our example, the asymptotic thresholds are slightly too large for predictions of $A(t)$ and $B(t)$ which makes the asymptotic confidence intervals conservative.
\label{fig:thresholdComparison}}
\end{figure}
A suitable parameter transformation makes the prediction profile likelihood equivalent to the parameter profile likelihood.
Therefore, the following discussion holds for both, for parameter and for prediction confidence intervals.

In general, fitting a model to experimental data reduces the residual sum of squares $\text{RSS}$.
In the asymptotic case, i.e.~for a large number of data points, it can be shown that the decrease of $\text{RSS}$ due to 
fitting one parameter is chi-square distributed with one degree of freedom.
This result also holds exactly in the non-asymptotic case for linear parameters.
This outcome is utilized to define the asymptotic threshold for profile likelihood confidence intervals \cite{Raue2009}.

For nonlinear parameters, the distribution of the decrease of the residual sum of squares by the parameter estimation procedure, has not yet been derived for the general setting. 
However, since the profile likelihood based confidence intervals are independent on bijective transformations of the parameter space \cite{cox94}, the assumption also holds if there is such a transformation, which makes the parameter of interest linear at least within its confidence interval.
Such a transformation only has to exist, it is not required to derive it analytically.

A situation where such a transformation does not exist occurs if the nonlinearity yields a non-monotone dependency of the profile likelihood, i.e.~there are several local minima in the confidence interval.
In this case, there is a larger decrease of the residual sum of squares and the standard threshold yields conservative results,
i.e.~the calculated confidence intervals are too large for the desired confidence level $\alpha$.

In Fig.~\ref{fig:checkThreshold}, a procedure is presented for checking the standard threshold.
It is a Monte-Carlo analysis of the impact of the nonlinear constraint used to calculate the prediction profile likelihood on the magnitude of overfitting.
In Fig.~\ref{fig:thresholdComparison}, the asymptotic thresholds corresponding to $\alpha = 0.05,0.1,\dots,0.95,0.99$, i.e.~the quantiles of the $\chi^2_1$ distribution, are compared with the empirically calculated thresholds 
for several prediction scenarios.
For the model 
\begin{equation}
     A\: \rightarrow \:B\: \rightarrow \:C
\end{equation}
the asymptotic thresholds are slightly too large for predicting $A(t)$ and $B(t)$ 
on the basis of measurements of $C(t)$. 
This makes the asymptotic confidence intervals conservative.
The impact on the coverage is discussed in the following section.

\subsection{Coverage}
\begin{figure}
\begin{center}
\includegraphics[width=\linewidth]{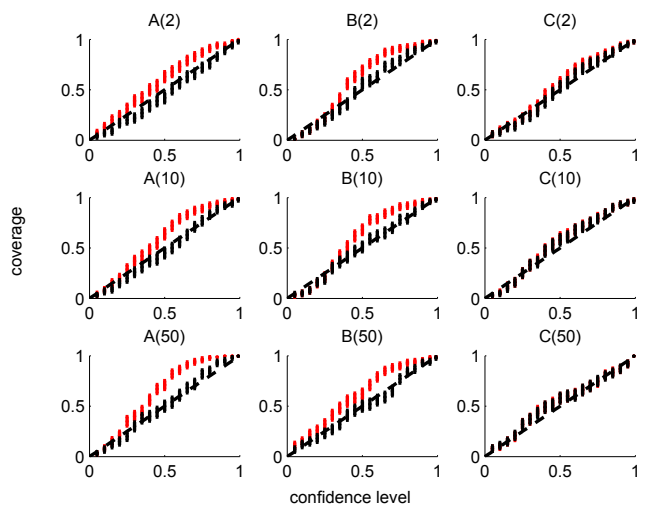}
\end{center}
\caption{Coverage of the prediction confidence intervals for the consecutive model.
The horizontal axis is the confidence level $\alpha=\{0.05,0.1,\dots, 0.95, 0.99\}$ which constitutes
the desired coverage of the confidence intervals.
The vertical axis is the realized coverage obtained for 100 data realizations. 
The red error bars are the result obtained for the asymptotic threshold which yield conservative outcomes for predictions of $A(t)$ and $B(t)$.
The black error bars indicate the results for the Monte-Carlo thresholds which shows almost perfect agreement with the confidence level in all prediction scenarios.
\label{fig:coverage}}
\end{figure}
The coverage 
\begin{equation}
	C = \text{Prob}( F(D_{\text{pred}},\theta_{\text{true}}) \in \text{PCI}_{\alpha}(D|y) ) 
\end{equation}
is the probability that the 
$\text{PCI}_{\alpha}(z|y)$ contains the true value $F(D_{\text{pred}},\theta_{\text{true}})$.
A desired property of any confidence interval is that the coverage coincides with the confidence level $\alpha$.

Fig.~\ref{fig:coverage} shows the estimated coverage of the prediction confidence intervals calculated for nine different prediction scenarios.
In these scenarios $A(2),$ $B(2),$ $C(2),$ $A(10),$ $B(10),$ $C(10),$ $A(50),$ $B(50),$ $C(50)$ have been predicted, 
i.e.~all three dynamic variables are predicted for an early, an intermediate, and a late point in time.
For this analysis, a hundred noise realizations have been analyzed.
The error bars plotted in this figure are bootstrap confidence intervals of the mean coverage.

The coverage obtained for the asymptotic threshold (red) tends to be conservative, i.e.~the true model response is inside the confidence interval more frequently as specified by the confidence level $\alpha$.
This means that there are more false negatives than intended which does not constitute a serious problem in terms of validity of conclusions.
In contrast, an anti-conservative coverage would constitute an issue because an increased false positive rate could lead to invalid reasoning.

The coverage obtained by the adjusted thresholds obtained by the Monte-Carlo algorithm shown in Fig.~\ref{fig:checkThreshold} are displayed by the black error bars in Fig.~\ref{fig:coverage}.
Here, the coverage coincides with the confidence level which confirms the validity respective the prediction profile likelihood based confidence intervals.

\subsection{Comparison of PCI and VCI}
\begin{figure}
\begin{center}
	\includegraphics[width=\linewidth]{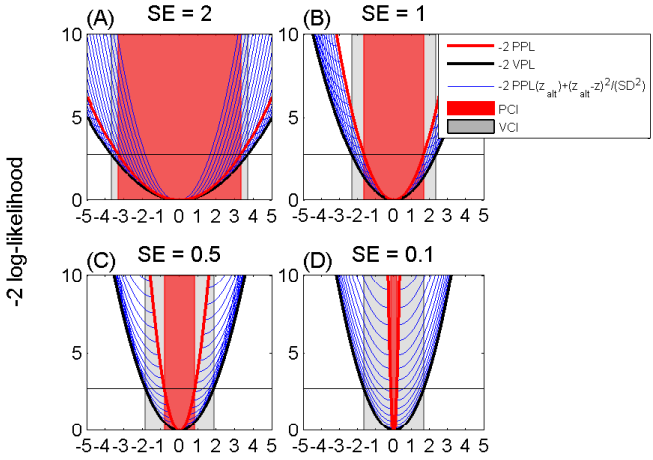}
\end{center}
\caption{Comparison of prediction and validation confidence intervals.
Panel (A) shows a prediction profile likelihood (red line) with a rather flat shape.
Here, the curvature of the prediction profile likelihood corresponds to a prediction standard error $\text{SE}=2$.
In this case, the prediction confidence intervals are large (red shaded) and the increase of the validation confidence intervals (gray) is smaller than indicated by the validation data error $\text{SD}$. 
If the data is more informative, i.e.~$\text{SE}$ decreases (panels B-D), the slope of prediction profile likelihood increases yielding larger difference between the $\text{PCI}$ and $\text{VCI}$.
\label{fig:pplVplIllu}}
\end{figure}
The validation profile likelihood satisfies
\begin{equation}
	\text{VPL}^{\text{SD}}(z|y) \le  \text{PPL}(z_{\text{alt}}|y) + \frac 1 2 \frac{(z_{\text{alt}}-z)^2}{\text{SD}^2}  \:\:\,\:\forall z_{\text{alt}} \label{eq:ineq}
\end{equation}
i.e.~the $\text{VPL}^{\text{SD}}(z|y)$ is smaller than the right hand side of the inequality for any alternative predicted value $z_{\text{alt}}$.
On the one hand, the inequality can be utilized to interpret a difference between the respective confidence intervals.
Furthermore, the equation can be utilized for consistency checks, e.g.~to prove the numerically calculated $\text{VPL}$ and $\text{PPL}$.
Small differences between the size of the $\text{VCI}^{\text{SD}}$ and $\text{PCI}$ indicate a flat prediction profile likelihood close to the threshold whereas deviations of the confidence intervals in the order of magnitude of $SD$ occur if the PPL has a large slope.
This aspect is illustrated in the following.

For illustration purpose, a quadratic prediction profile likelihood
\begin{equation}
	-2\,\text{PPL}(z) = \frac{z^2}{\text{SE}^2}
\end{equation}
with $\text{SE} \in \{0.1,\,0.5,\,1,\,2\}$ has been assumed.
These four settings are shown in Fig.~\ref{fig:pplVplIllu}.
The prediction profile likelihood is shown as a red line.
For several $z_{\text{alt}}$, the quadratic term in [$\ref{eq:ineq}$] is plotted by blue curves attached to the PPL. 
The $\text{VPL}$ constitutes the infimum of these curves which in this special case
can be calculated analytically and is given by
\begin{equation}
	-2\,\text{VPL}^{\text{SD}}(z)=\frac{z^2 \text{SE}^2}{\left(\text{SD}^2+\text{SE}^2\right)^2} 
	+ \left(\frac{z}{\text{SD}}  - \frac{z\text{SE}^2}{\text{SD} \left(\text{SD}^2+\text{SE}^2\right)} \right)^2
\:\:.
\end{equation}

Panel (A) shows the comparison for $\text{SE}>\text{SD}$. 
In this case, the boundaries of the VCI and the PCI differ only by a value around 0.38.
In Panel (B), $\text{SE}$ is chosen equal to $\text{SD}$. 
In Panels (C) and (D) $\text{SE}$ is further decreased. 
This corresponds to more informative data for predicting the exact value of $z$.
In these cases, the optimum of the PPL is narrow in comparison to validation data error $\text{SD}$.
Then, during fitting the model, a mismatch $z-z^*$ is predominantly explained by the observation error of the validation data point.
The difference of the boundaries of the confidence intervals increase and approach the 10\% quantile of the Gaussian distribution, i.e.~a value $\text{icdf}(N(0,\text{SD}=1),.95) =1.64$ which is the one-sided 5\% confidence interval for a validation data point for a constant model prediction, i.e.~for $\text{SE}\rightarrow 0$.
 
\subsection{Prior information \label{sec:prior}}
\begin{figure*}[h]
\begin{center}
\includegraphics[width=0.3\linewidth]{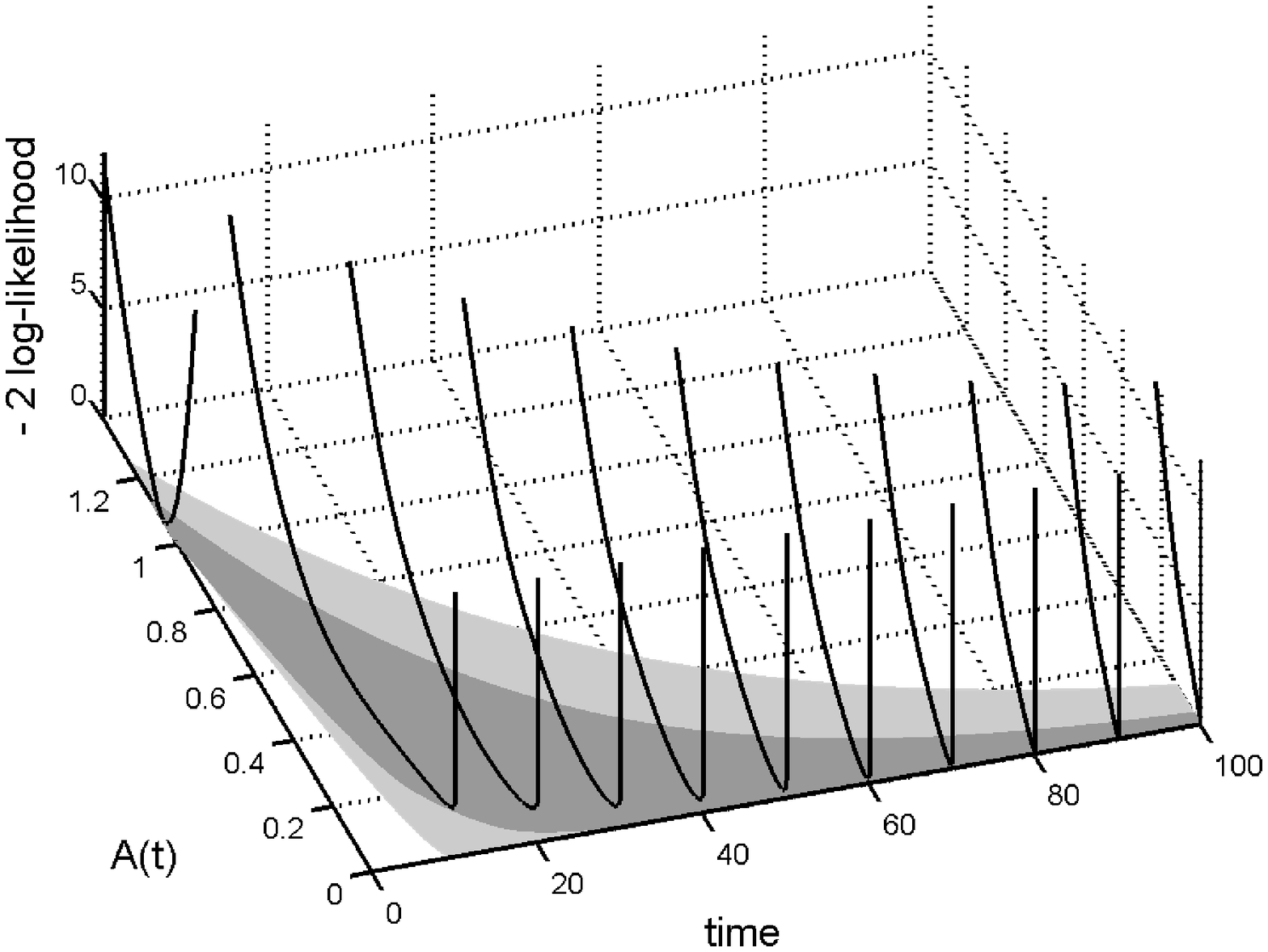}
\includegraphics[width=0.3\linewidth]{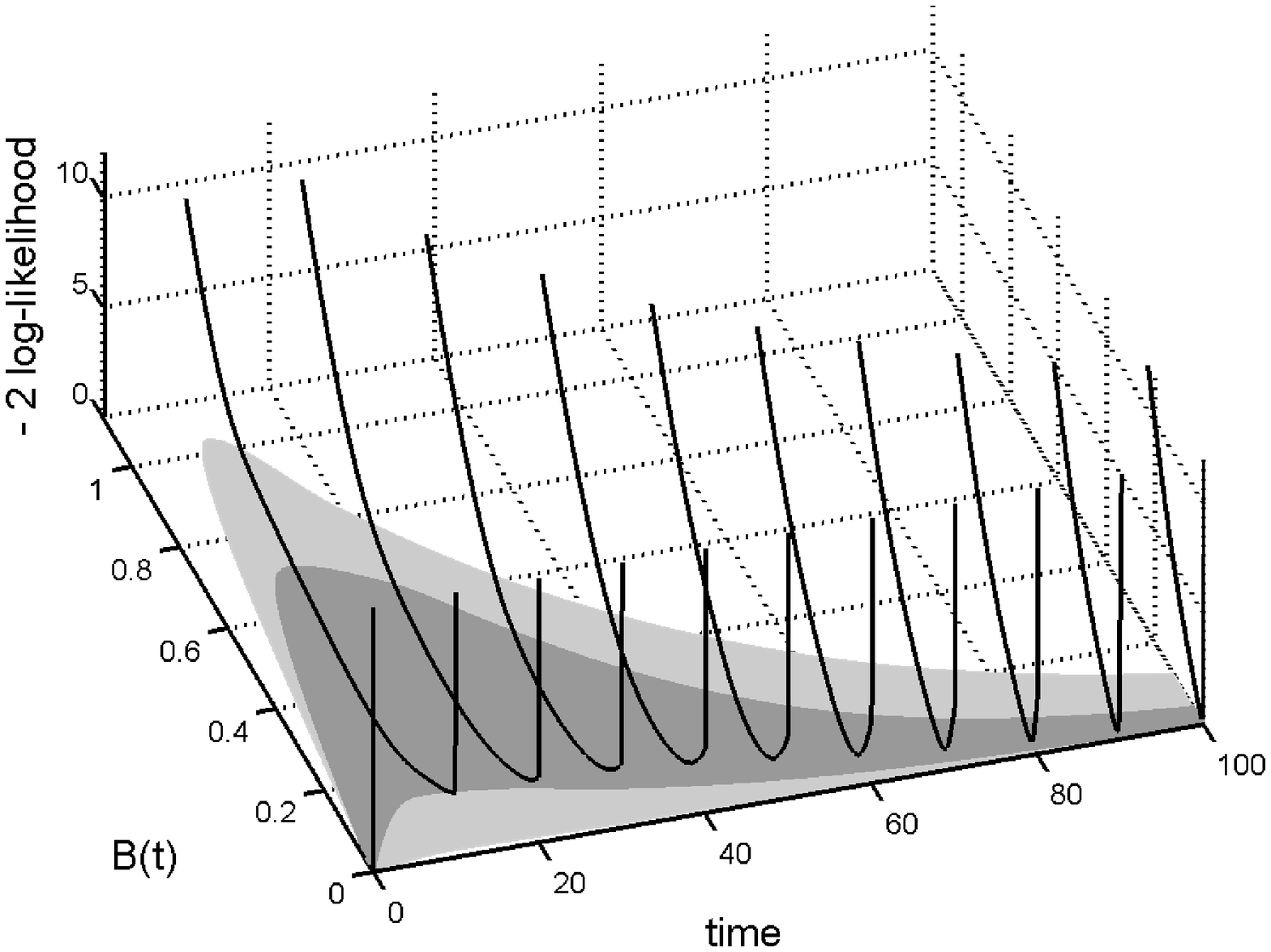}
\includegraphics[width=0.3\linewidth]{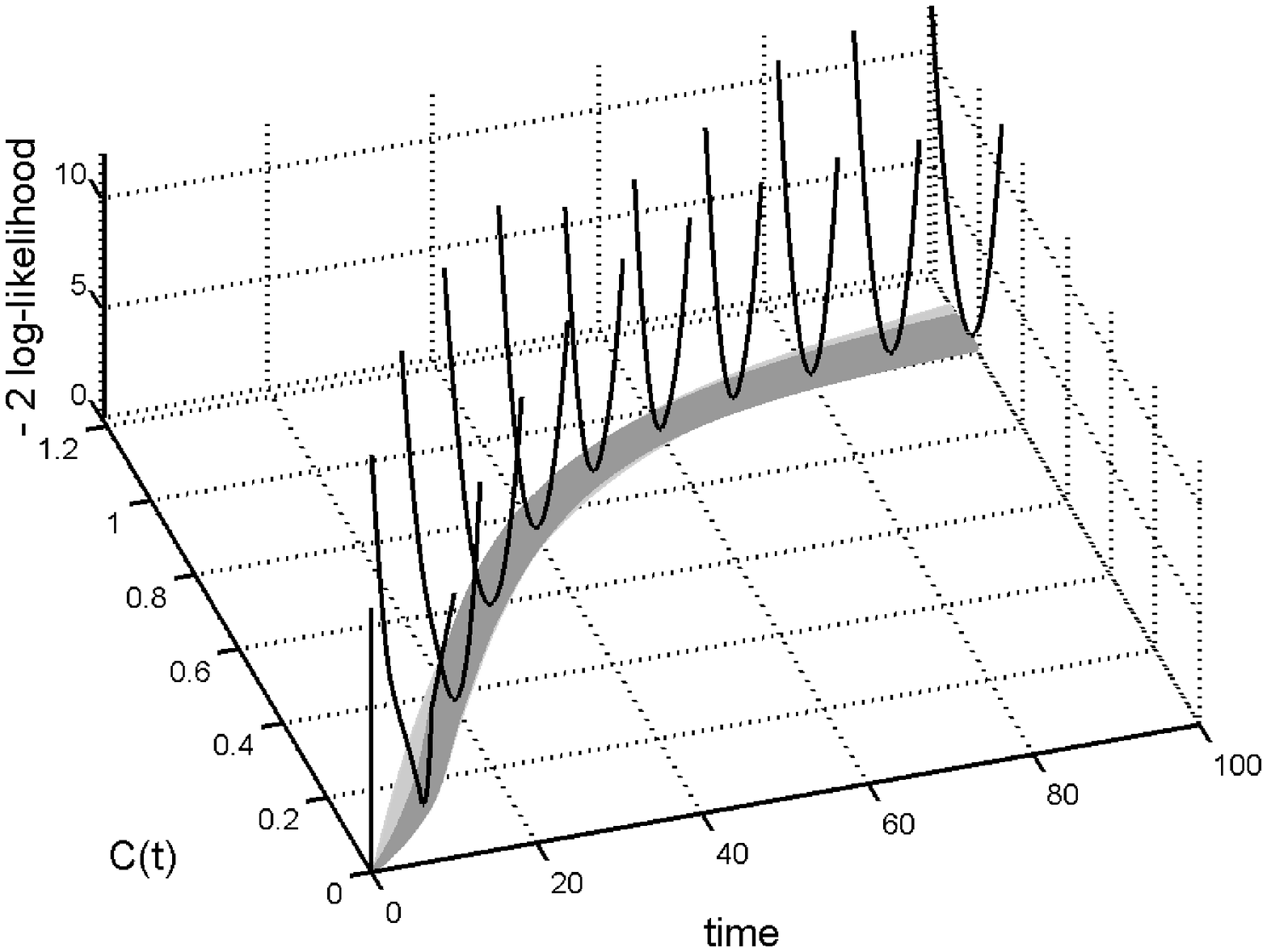}
\end{center}
\caption{The curves in vertical direction are the prediction profile likelihood functions for $A(t)$ (left panel), $B(t)$ (middle), and $C(t)$ (right panel) if a log-normal prior for $\theta_3$ is assumed.
The respective 90\% confidence intervals are plotted in dark gray.
The light gray regions indicate the 90\% confidence intervals if the parameter $\theta_3$ is estimated without prior information.
\label{fig:pplPrior}
}
\end{figure*}
If prior information about parameters is available, e.g.~a prior distribution $\pi(\theta)$, maximum likelihood estimation is replaced by maximum a-posteriori (MAP) estimation 
\begin{eqnarray}
	\hat \theta_{\text{MAP}} &=& \arg \max_{\theta} \rho(y|\theta) \pi(\theta)  \\
		&=& \arg \max_{\theta} \left( \text{LL}(y|\theta) + \log(\pi(\theta))  \right) \:\:,
\end{eqnarray}
i.e.~the parameters are estimated by maximizing the a-posterior probability of the data and the parameter estimates.
For most common priors, MAP estimation can be performed by MLE using a penalized likelihood.
As an example, a log-normal prior for a parameter component $\theta'$ yields
\begin{equation} \label{eq:LLpen}
	\text{LL}_{\text{prior}} = \text{LL} - \frac 1 2 \frac{\left( \log(\theta') - \langle \theta' \rangle \right)^2}{\text{Var}(\theta')} + const\:\:.
\end{equation}
To incorporate prior knowledge, the presented prediction profile likelihood approach 
has to be generalized to MAP estimation and the penalized likelihood [$\ref{eq:LLpen}$] is used instead of the standard log-likelihood $\text{LL}$.

To illustrate the incorporation of prior knowledge for parameter values,
the initial concentration $A(0)=\theta_3$ is assumed to be drawn from a log-normal distribution  
\begin{equation}
	\theta_3 \sim logN(0,1) \label{eq:logN}
\end{equation} with expectation $\langle \log(\theta_3) \rangle = 0$ and variance $Var(\log(\theta_3))=0.1$.
For parameter estimation, this is accounted for by using the penalized likelihood [$\ref{eq:LLpen}$], 
i.e.~by adding an additional term to the residual sum of squares.

As in the example in the main text, the calculation of the prediction and validation confidence intervals has been repeated for $t=0,10,\dots,100$ and all three dynamic states $A(t)$, $B(t)$, $C(t)$.
In this example, the true value of $A(0)\equiv \theta_3$ has been drawn according to the prior from the log-normal distribution [$\ref{eq:logN}$]. 

Fig.~\ref{fig:pplPrior} shows the prediction profile likelihood functions as curves in vertical direction as well as the respective 90\% prediction confidence intervals as dark gray shaded regions.
The prediction confidence intervals plotted in light color are obtained if $\theta_3$ is estimated without prior information.
Because $C$ is the measured compound in our example, the prediction confidence intervals for $C$ are much smaller than for $A$ and $B$.
However, also $A$ and $B$ yield bounded prediction confidence intervals which can be interpreted as {\it observability} of these dynamic states.
Omitting the prior information yields larger prediction confidence intervals, 
especially for the unobserved states $A(t)$ and $B(t)$.

\subsection{Validation profile likelihood for the consecutive reaction model}
\begin{figure*}[h]
\begin{center}
\includegraphics[width=0.3\linewidth]{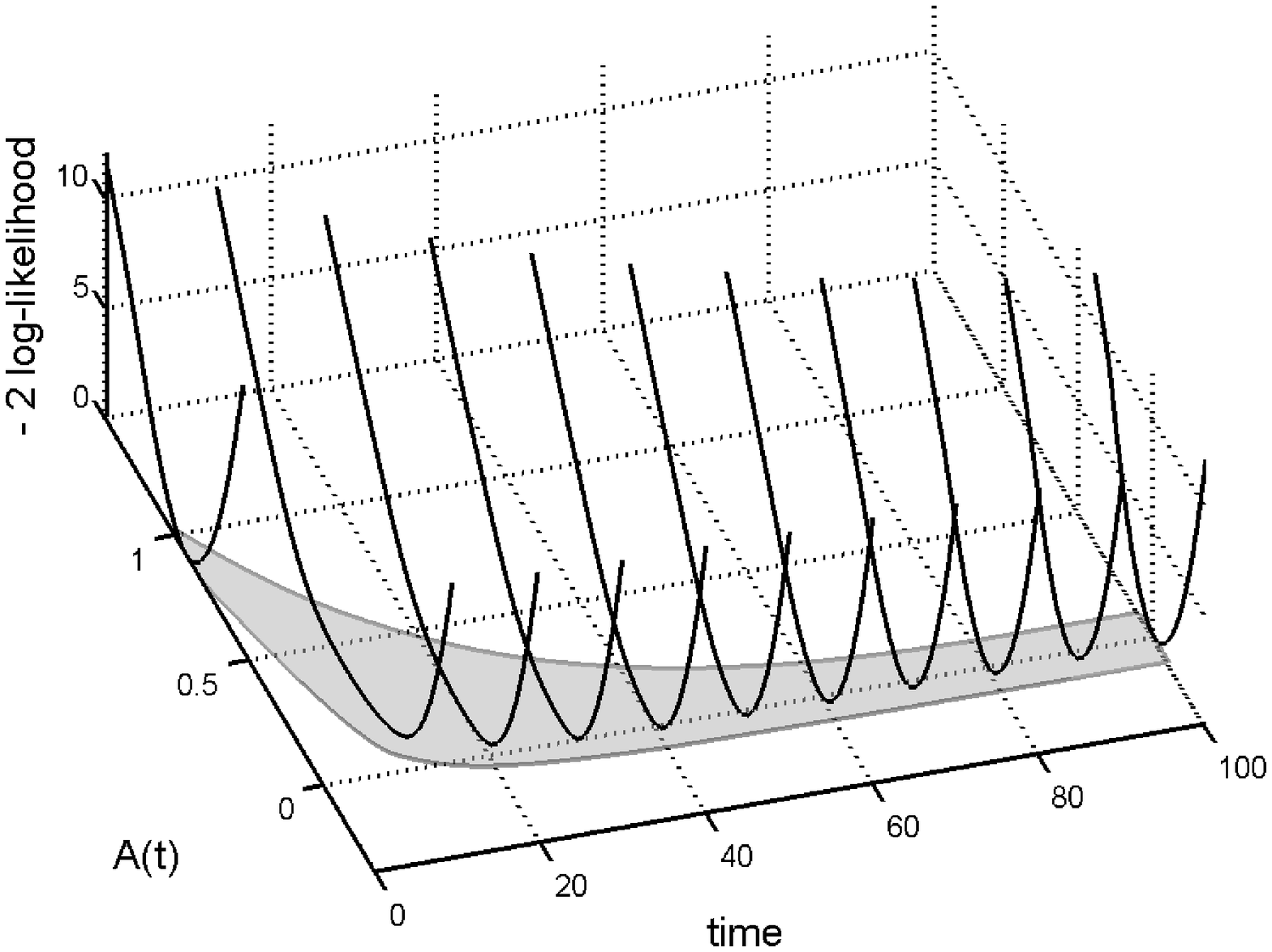}
\includegraphics[width=0.3\linewidth]{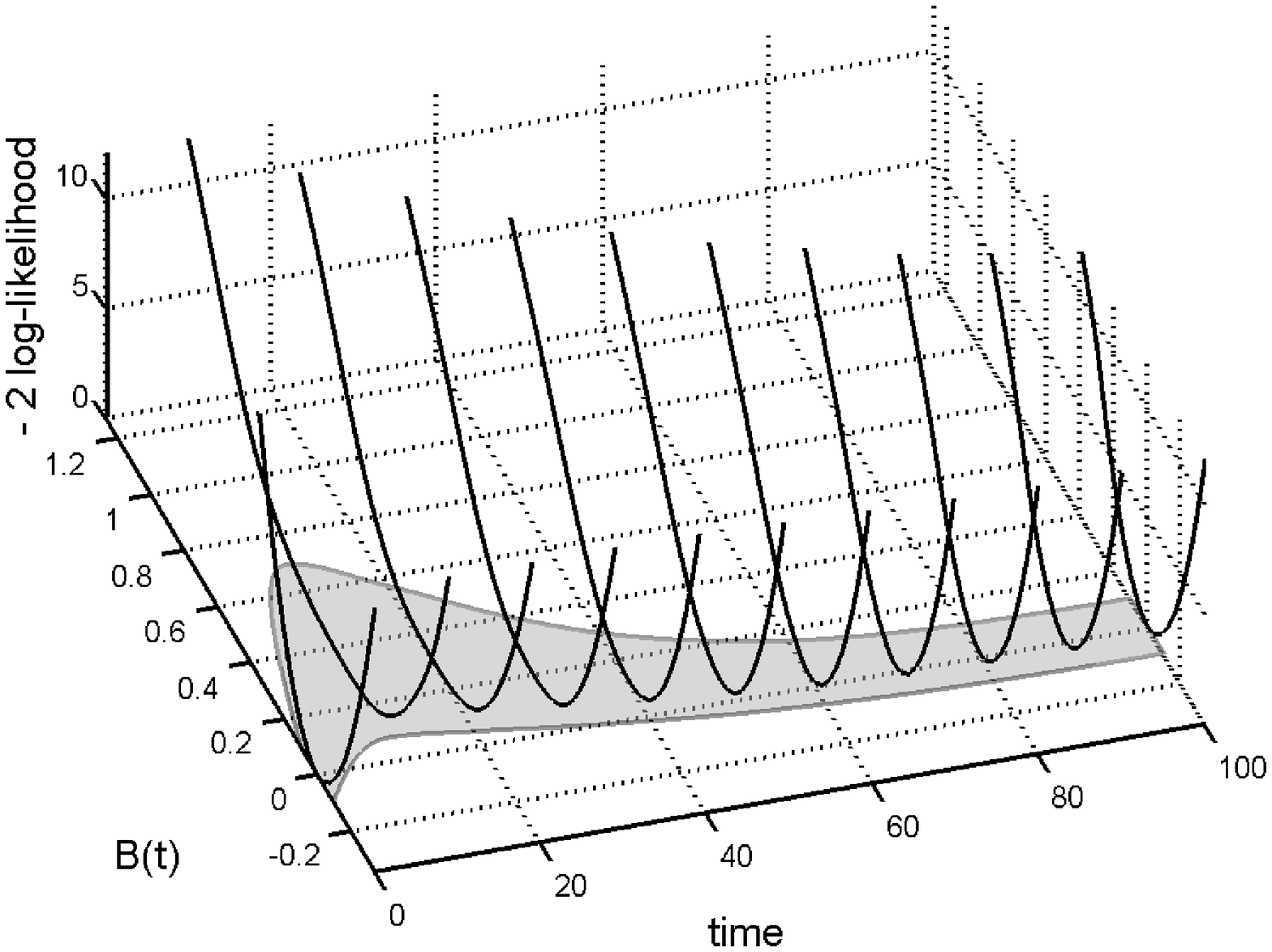}
\includegraphics[width=0.3\linewidth]{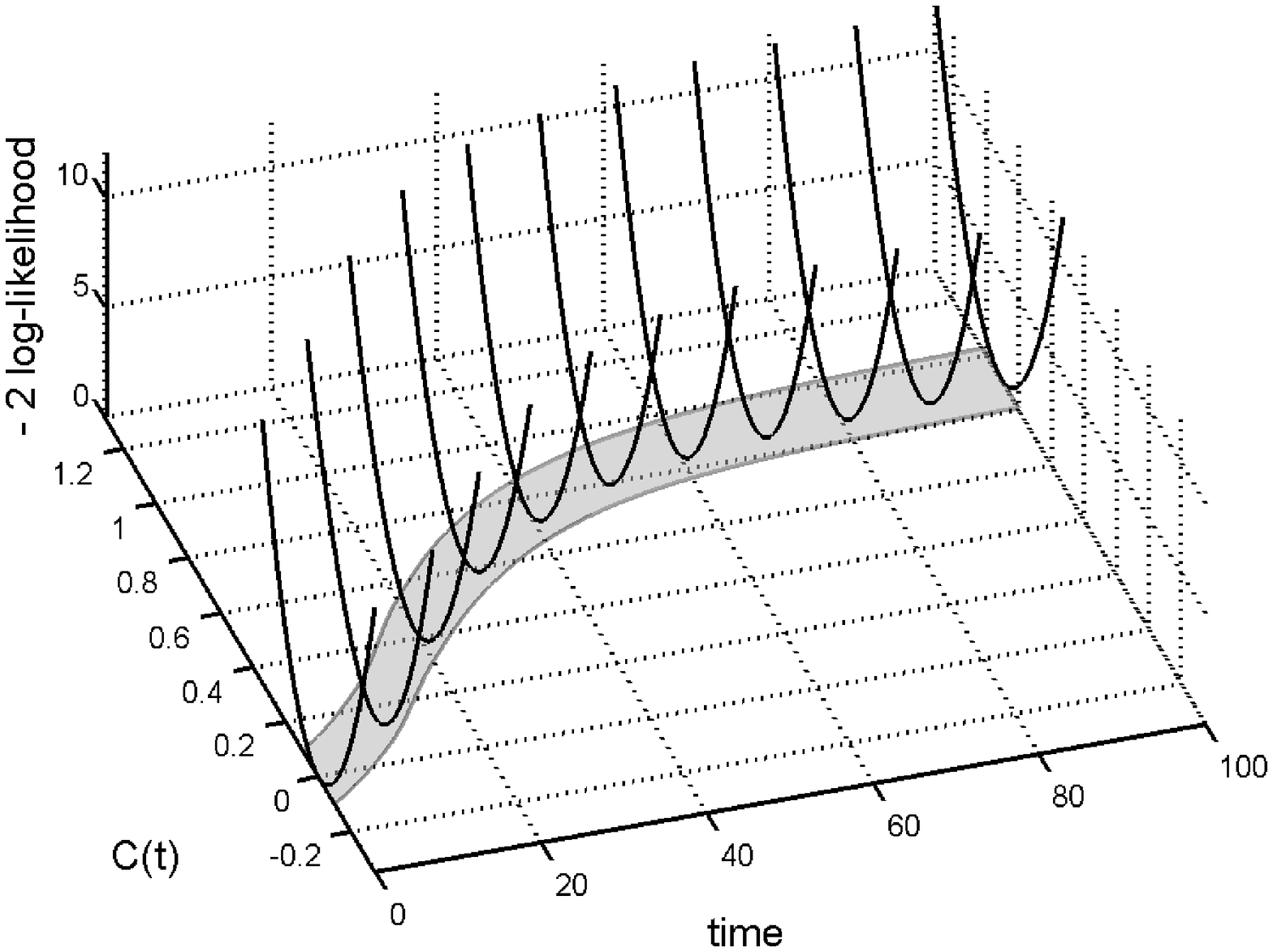}
\end{center}
\caption{The left panel shows the validation confidence intervals for the unobserved state $A(t)$.
The validation profile likelihood functions are plotted as curves in vertical direction.
For the plotting the confidence intervals along the time axis, the $\text{VCIs}$ have been interconnected by cubic piecewise interpolation.
Validation confidence intervals and validation profile likelihood functions for the intermediate unobserved state $B(t)$ are shown in the middle and for $C(t)$ in the right panel.
\label{fig:cons3Dvpl}
}
\end{figure*}
In the main text, prediction confidence intervals have been shown for the consecutive reaction model.
Fig.~\ref{fig:cons3Dvpl} shows the corresponding validation profile and the respective validation confidence intervals for the same noise realization
for all dynamic variables $A(t)$, $B(t)$, and $C(t)$.
Validation confidence intervals account for the measurement noise in a validation experiment.
Therefore, they are larger than the prediction confidence intervals shown in the main text in Fig.~1, panels (c)-(e).

Because Gaussian noise $\varepsilon\sim N(\mu,SD^2)$ has been assumed, the validation confidence intervals
covers negative values if the true model response $\mu=F(D_{\text{pred}}, \theta_{\text{true}})$ is close to zero.

\subsection{Observability of the long-term dynamics}
\begin{figure}
\begin{center}
	\includegraphics[width=\linewidth]{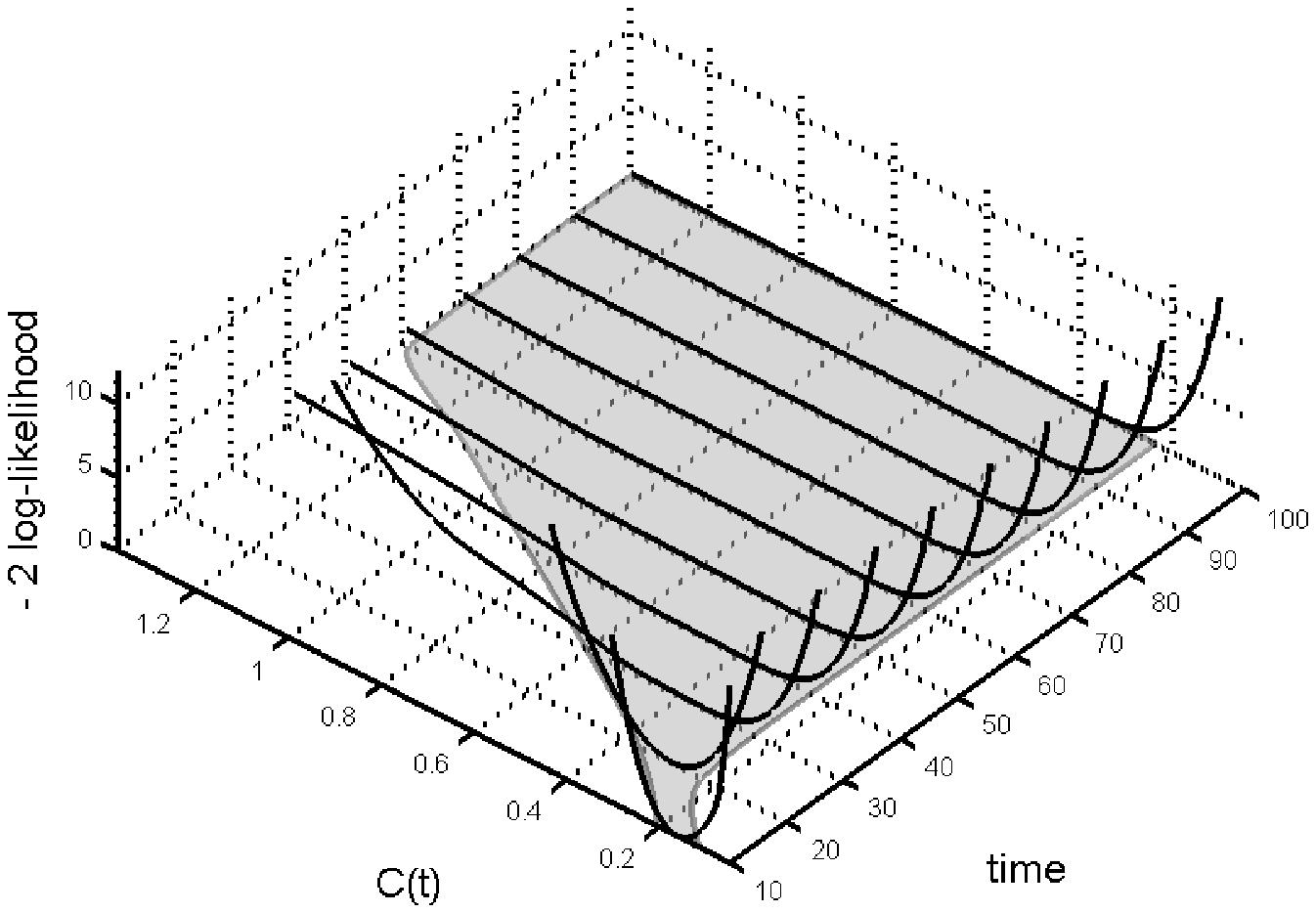}
\end{center}
\caption{Prediction confidence intervals for the extrapolation of $C(t)$ to time points much larger than the measurement times.
Because in this example the experimental design does not provide sufficient information about the steady state level of $C(t)$,
the prediction confidence intervals diverge and the steady state $C(t)$ for $t\rightarrow \infty$ is non-observable.
\label{fig:nonobsC}}
\end{figure}
In the main text, it has been discussed how to exploit the prediction profile likelihood for observability analyses.
For the two step model 
\begin{equation}
     A\: \overset{\theta_1}{\rightarrow} \:B\: \overset{\theta_2}{\rightarrow} \:C
\end{equation}
it has been shown that measurements of the compound $C$ which sample only the transient increase and therefore does not provide information about the steady state level lead to non-observability of $A(t)$, and $B(t)$ for $t>0$.
In addition to that result, Fig.~\ref{fig:nonobsC} shows prediction confidence intervals for $C(t)$ for times much larger than the measurement times $t=0,2,\dots,20$.
$C(t)$ becomes practically non-observable for times which are much larger than the time sampling interval.

\subsection{Characteristics of the MAP kinase model}
\begin{figure}
\begin{center}
	\includegraphics[width=\linewidth]{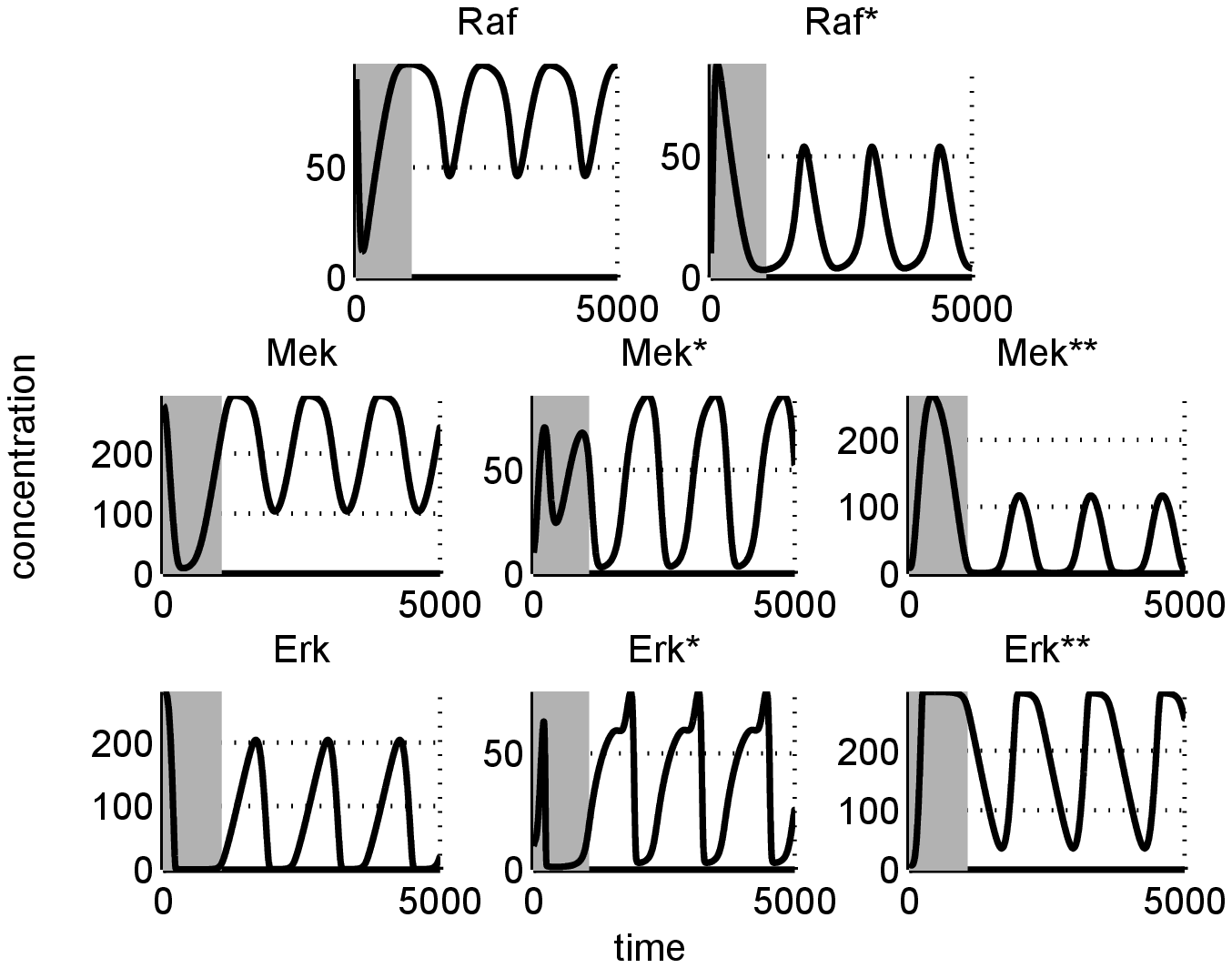}
\end{center}
\caption{The long-term dynamics of the MAP kinase model shows regular oscillations.
In our analysis, the first 1000 seconds, as highlighted by gray background color, have been analyzed as a compromise between a transient and oscillatory dynamics.\label{fig:kholo_long}}
\end{figure}
\begin{table*}
\centering
\begin{tabular}{clcccc}
symbol&description&value &lower boundary&upper boundary & units\\
\hline
$V_1$&	max. enzyme rate &	2.5&		1e-8  & 1e6 & nM s$^{-1}$\\
$n$ &	Hill coefficient of the feedback	&	1&		1  & 1 & 1\\
$K_I$&	Michaelis constant &	9&		1e-8  & 1e6 & nM\\
$K_1$&	Michaelis constant &	10&		1e-8  & 1e6  & nM\\
$V_2$&	max. enzyme rate &	0.25&		1e-8  & 1e6  & nM s$^{-1}$\\
$K_2$&  Michaelis constant 	&	8&		1e-8  & 1e6  & nM\\
$k_3$&	catalytic rate constant &	0.025&		1e-8  & 1e6  & nM s$^{-1}$\\
$K_3$&	Michaelis constant &	15&		1e-8  & 1e6  & nM\\
$k_4$&	catalytic rate constant &	0.025&		1e-8  & 1e6  & nM s$^{-1}$\\
$K_4$&	Michaelis constant &	15&		1e-8  & 1e6  & nM\\  
$V_5$&	max. enzyme rate &	0.75&		1e-8  & 1e6  & nM s$^{-1}$\\
$K_5$&	Michaelis constant &	15&		1e-8  & 1e6  & nM\\
$V_6$&	max. enzyme rate &	0.75&		1e-8  & 1e6  & nM s$^{-1}$\\
$K_6$&	Michaelis constant &	15&		1e-8  & 1e6  & nM\\  
$k_7$&	catalytic rate constant &	0.025&		1e-8  & 1e6  & nM s$^{-1}$\\
$K_7$&	Michaelis constant &	15&		1e-8  & 1e6  & nM\\
$k_8$&	catalytic rate constant &	0.025&		1e-8  & 1e6  & nM s$^{-1}$\\
$K_8$&	Michaelis constant &	15&		1e-8  & 1e6  & nM\\  
$V_9$&	max. enzyme rate &	0.5&		1e-8  & 1e6  & nM s$^{-1}$\\
$K_9$&	Michaelis constant &	15&		1e-8  & 1e6  & nM\\
$V_{10}$	&max. enzyme rate 	&	0.5&		1e-8  & 1e6  & nM s$^{-1}$\\
$K_{10}$	&max. enzyme rate 	& 15&	1e-8  & 1e6  & nM\\
\hline
\end{tabular}
\caption{Parameters of the MAP kinase model as published in \cite{Kholodenko2000}.\label{tab:kholodenko_parameters}}
\end{table*}

To demonstrate the applicability of our approach in a realistic setting, 
the published model of MAP kinase signaling \cite{Kholodenko2000} has been utilized to illustrate the calculation of
prediction and validation confidence intervals.

Fig.~\ref{fig:kholo_long} shows the long-term dynamics of this model, i.e.~the oscillations.
In our analysis only the initial phase, i.e.~the first 1000 seconds have been considered.
This time interval is characterized by strong nonlinearity of the model response with respect to the parameters
and constitutes a compromise setting between a transient and an oscillatory dynamics.

Tab.~1 summarizes the model parameters as they have been published in \cite{Kholodenko2000}.
The Hill coefficient $n$ is assumed to be equal to one.
For the observational noise of the validation data, the same noise level as for the experimental data has been assumed, 
i.e.~$\sigma=\text{SD}=10$.

\subsection{Experimental design conclusions}
\begin{figure}
\begin{center}
	\includegraphics[width=\linewidth]{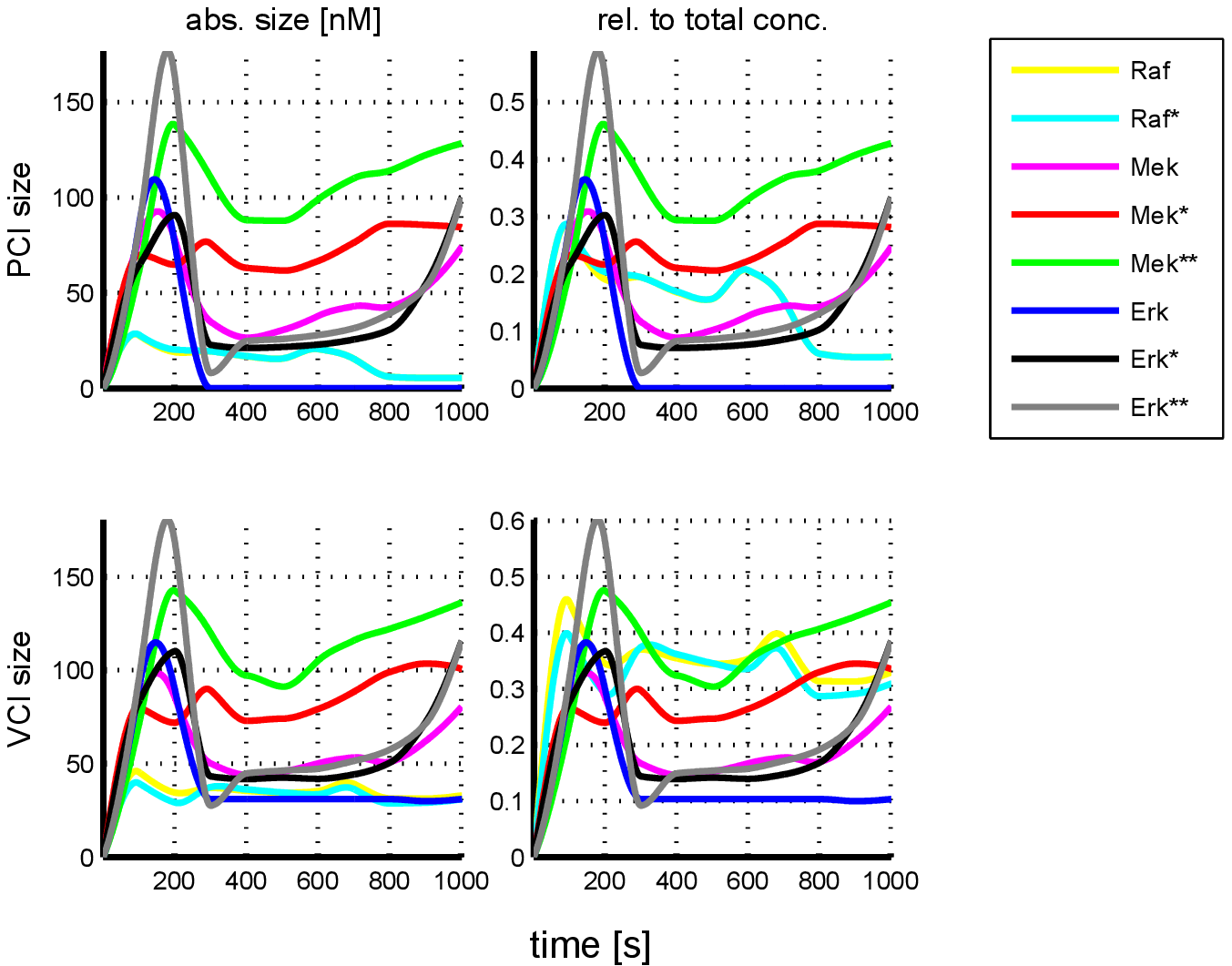}
\end{center}
\caption{Size of the prediction confidence intervals for the dynamic states of the MAP kinase model.
The left panels show the size of the confidence intervals in absolute units. 
In the panels on the right, the size is plotted relative to the total concentration of a protein.
The upper row shows the results for the prediction confidence intervals, the lower row for the validation confidence intervals.
\label{fig:pciSize}}
\end{figure}
The size of prediction confidence interval can be utilized to figure out informative experimental designs. 
If the information of a single data points is intended to be evaluated, then the validation confidence intervals are appropriate.
If many experimental replicates are feasible, the average observation will have a small standard error and then prediction confidence intervals can be used to assess a design.

Fig.~\ref{fig:pciSize} shows the size of the 90\% prediction confidence intervals (upper row), i.e.~the difference between the upper and lower boundary, and the size of the validation confidence intervals (lower row) along the time axis.
The size is plotted in absolute concentrations (left panels) and relative to the total amount of the protein (right panels).

Independently from the way of the assessment, Erk (blue lines) yields the smallest prediction and validation confidence intervals for $300<t\le1000$.
Therefore, measurements of Erk in this time interval constitute very informative experimental designs for testing the model.
As discussed in the main text, such a setting which is appropriate for validating the whole model is not informative for improving the model parameters.
For such a purpose, new experimental data has to be generated for designs 
in which the model behavior is not yet precisely specified, i.e.~for a setting with large prediction or validation confidence intervals.
In these terms, Erk$^{**}$ (gray lines) is most informative in absolute units between 100 and 200 seconds.
Also absolute measurements of Mek$^{*}$ (red lines) and Mek$^{**}$ (green lines) along the whole time axis are informative.

If only the amount of a phosphorylated form relative to the total concentration of the protein is experimentally accessible, then the panels on the right should be evaluated to assess the power of a design. 
In our example, the outcome for the prediction profile likelihood is very similar to the results obtained for absolute concentrations. 
Again, Erk for $t<200$ as well as Mek$^{*}$ and Mek$^{**}$ are most informative.
For the validation confidence intervals, Raf and Raf$^*$ appear as informative because the total concentration of Raf is three times smaller than the total amount of Mek and Erk.

\subsection{Implementation}
The cvodes package from the Suite of Nonlinear Differential/Algebraic Equation Solvers (SUNDIALS) \cite{Hindmarsh2005}
has been used for the numerical integration of the ODEs and the sensitivity equations. MATLAB's fmincon optimizer was used to estimate the parameters. The gradient and the Hessian of the objective function have been provided for the optimizer using the sensitivity equations \cite{Leis:1988dl}
and the approximation
\begin{equation}
	\frac{\partial^2}{\partial \theta_j \partial \theta_k} \text{LL} \approx \sum_i \frac 1 {\sigma_i^2}\frac{\partial F_i}{\partial \theta_j} \frac{\partial F_i}{\partial \theta_k}
\end{equation}
of the second derivatives \cite{Press92}. 
Within a single optimization procedure, the parameters have been alternatingly optimized on the logarithmic scale as well as the common linear scale until the optimizer converged to a common value on both scales.

For the calculation of the validation profile likelihood, 30 test predictions $z$ within the reasonable range of predictions given by the model structure and $\text{SD}$ have been evaluated to obtain an initial guess of the likelihood shape.
Then the grid is iteratively refined and/or enlarged until a smooth validation likelihood covering the whole confidence interval is obtained and local minima have been removed.
For a single profile likelihood, around $10^2$ to $10^3$ optimizations were required.

For the prediction profile likelihood the initial guess is obtained from the VPL by equation [$17$] in the main text. 
The gaps in this guess are then filled by nonlinear constrained optimization.
If the constraint optimization procedure did not converge, the validation data error $\text{SD}$ has been iteratively decreased by factors $10^0,10^{-1},\dots,10^{-5}$.


\end{document}